\documentclass[10pt,a4paper]{article}

\usepackage{fullpage}
\usepackage[utf8x]{inputenc}
\usepackage{graphicx}
\usepackage{amsmath}
\usepackage{amssymb}
\usepackage{subfigure}

\newcommand{\ie}{\textit{i.e.}\ }
\newcommand{\eg}{\textit{e.g.}\ }
\newcommand{\cf}{\textit{cf.}\ }
\newcommand{\pdown}{\raisebox{-1ex}{.}}
\newcommand{\eq}{0}
\newcommand{\fo}{\text{f.o.}}
\newcommand{\BJ}{\mathrm{BJ}}

\begin{document}
\title{\textbf{Iterative Fluiddynamics}}

\author{F. Wunderlich and B. K{\"a}mpfer\\[2ex]
\textit{Helmholtz-Zentrum Dresden-Rossendorf D-01314 Dresden, Germany}}
\date{\today}
\maketitle
\textbf{Abstract.}\ \ 
   An iterative scheme is presented to solve analytically the relativistic fluid dynamics equations. The scheme is 
   applied to longitudinal expansion, transversal symmetric and transversal asymmetric (triaxial)
   expansion as well. Within this scheme it is possible to describe the dynamics of a strongly coupled 
   (\ie conformal) medium for parameters referring to heavy-ion collisions at LHC.
\\[2ex]

\section{Introduction}
\label{intro}
Fluid dynamics became a standard tool for modeling the expansion dynamics of the strongly coupled quark-gluon plasma (sQGP)
produced in ultrarelativistic heavy-ion collisions. For given initial conditions, \eg inferred from transport models or 
Glauber type 
Monte Carlo simulations, the sQGP's evolution is followed in space and time until the density becomes so low that a description by 
hadronic degrees of freedom is appropriate and the late history until and including the freeze-out can be described 
by a hadronic transport model. In fact, various observables, such as transverse momentum spectra, differential elliptic flow 
and higher flow modes as well, can be successfully explained for RHIC and LHC beam energies \cite{Kolb:2001qz,Shen:2011eg}.
The initial conditions for the subsequent fluid dynamical evolution might be smooth, \ie emerging from an average 
over many events, or might be fluctuating, \ie corresponding to an event-by-event evolution \cite{Aguiar:2001ac}.
The influence of initial state fluctuations on final state observables currently receives much attention 
(for a selection of recent works \cf \cite{Qiu:2011iv})
which to a large extent is motivated by a detailed understanding of the particle distributions and correlations beyond 
the elliptic flow mode as well as transport properties of the expanding medium.
\par
At the heart of the fluid dynamical description are the equation of state, accessible in lattice QCD calculations 
\cite{Borsanyi:2011zzb,Aoki:2005vt,Huovinen:2011xc,Ejiri:2009hq},
and transport coefficients, being less directly accessible by lattice QCD \cite{Karsch:1986cq,Aarts:2002vx,CasalderreySolana:2011us}
due to their real-time nature. 
The AdS/CFT correspondence, however, allows estimates of the transport coefficients 
(cf. \cite{CasalderreySolana:2011us,Kovtun:2004de,Buchel:2007mf}) and further properties of the matter in the strong coupling 
limit, thus complementing perturbative weak-coupling approaches. 
In such a way, quantitative results for the properties of the sQGP can be obtained. 
Nevertheless it remains not yet clear, to which extent AdS/CFT results can be adopted.
Intriguing questions concern the details of the confinement transition region, position of the conjectured (critical) 
end point in the phase diagram and the temperature dependence of the transport coefficients
in various regions of the phase diagram \cite{Friman:2011zz}.
\par
Relativistic fluid dynamics might be grouped into perfect fluid dynamics, \ie neglecting dissipative phenomena,
dissipative fluid dynamics, \eg as Landau-Lifschitz formulation \cite{Landau:1953gs,Belenkij:1956cd}, 
and extensions thereof such as the
Israel-Stewart type setup \cite{Israel:1979wp} or transient fluid dynamics \cite{Denicol:2012cn}. These fluid dynamical 
descriptions (\cf \cite{Van:2007pw,Van:2011yn} for further discussion of relativistic dissipative hydrodynamics) 
represent partial 
differential equations which  are usually solved numerically by employing suitable differencing schemes. Only in case of high 
symmetries of the flow, analytical solutions are at hand. Examples, relevant for heavy-ion collisions, are Bjorken's flow
 \cite{Bjorken:1982qr}, Gubser's flow \cite{Gubser:2010ze} and the flow patterns studied by Cs\"org\H o \textit{et al.}
\cite{Csorgo:2003rt,Csorgo:2003ry,Csorgo:2006ax}. 
(Further flow patterns refer to Landau type initial conditions,
\cf \cite{Csernai:1994xw}.) With respect to the important role of elliptic flow, however,
more realistic solutions are desirable. The elliptic flow analysis with account of longitudinal pressure gradients calls for
the time evolution of an initially triaxial matter distribution. This goal is envisaged in the present paper. We are going to utilize
a scheme for solving algebraically the fluid dynamical equations for arbitrary but smooth and analytically given
initial conditions. We test an iterative procedure which allows for solving fully analytically the fluid equations of
motion up to a given order in the expansion of powers of the time coordinate. Such iterative procedures have been established 
in the context of general relativity (\cf \cite{deHaro:2000xn}). In employing the AdS/CFT correspondence,
the iterative procedure is exploited, \eg  in \cite{Janik:2010we}, to solve the five-dimensional vacuum Einstein equations with
negative cosmological constant and to derive useful insights in the behavior of the four-dimensional finite-temperature,
strongly coupled medium in the conformal limit.
\par
Here, we outline such a solver for the four-dimensional fluid dynamics 
in section \ref{sec:Bas} and make it explicit for ideal fluid dynamics in section \ref{sec:IdFl}. Various applications up to the 
triaxial expansion dynamics including eccentricity as basis of the elliptic flow for conditions 
referring to heavy-ion collisions at LHC are presented in section \ref{sec:Examples}. We discuss the limitations
of the method in section \ref{sec:Limits} and draw conclusions in section \ref{sec:Conclusion}.

\section{Description of the iterative scheme}
\label{sec:Bas}
The basic idea of the iterative scheme to be described in the following paragraphs is to transform the differential 
equations of interest (in our case: local energy-momentum conservation) into an infinite set of algebraic relations
between the Taylor coefficients of the unknown functions, which can be solved iteratively. For achieving this, the unknown
functions need to be formally Taylor expanded and afterwards the series is plugged into the differential equation.
Comparing coefficients of the terms with the same power in the time-coordinate leads to the desired set of algebraic relations
which is solved iteratively up to a truncation order.
\par 
For this purpose, let us assume that the only fluid dynamical equations governing the dynamics of the sQGP are
\begin{equation}
   {T^{\mu\nu}}_{;\nu} \equiv \partial_\nu T^{\mu\nu} + \Gamma^\mu_{\alpha\nu}T^{\alpha\nu} + \Gamma^\nu_{\beta\nu}T^{\mu\beta}=0, 
                            \label{hyd:01}
\end{equation}
where the Christoffel symbols $\Gamma^\mu_{\alpha\beta}$ take into account non-Cartesian coordinates mapping out the four
dimensional Minkowski space-time. In the absence of conserved charges the symmetric energy-momentum tensor $T^{\mu\nu}$ 
is assumed to be related by a constitutive
equation to energy density $e$, pressure $p$, four-velocity $u^\mu$ and additionally to gradients of these quantities
in the form 
\begin{equation}
   T^{\mu\nu} = e u^\mu u^\nu + p \Delta^{\mu\nu} + \Pi^{\mu\nu},\label{hyd:02}
\end{equation}
with $\Delta^{\mu\nu}=u^\mu u^\nu - g^{\mu\nu}$ \footnote{In Minkowski space we use the "mostly minus" sign convention for the 
metric.}  being the projector orthogonal to $u^\mu$. The quantity $\Pi^{\mu\nu}$ uncovers the dissipative effects. 
An equation of state $e=e(p)$ and the normalization $u^\mu u_\mu= 1$ closes 
(\ref{hyd:01}) to a system of four evolution equations for four independent quantities, \eg $e$ and the spatial 
components $u^a$
of the four-velocity. The Landau Lifschitz condition $u_\mu\Pi^{\mu\nu}=0$ provides a unique definition of the velocity (\cf 
\cite{Van:2007pw,Tsumura:2012ss} for that issue). 
Cauchy initial conditions are specified by giving $e(\theta_0, \vec{\xi})$ and $u^a(\theta_0, \vec{\xi})$ on
the hypersurface $\theta=\theta_0$ for a suitable time coordinate $\theta$.
For the subsequent calculations we use the coordinates $\xi^\mu \sim (\Theta, \eta, \vec{x}_\perp)$ being related to 
Cartesian coordinates 
$(t, \vec{x})$ via $\Theta = \ln\frac{1}{\tau_\eq}\sqrt{t^2-{(x^1)}^2}$,
$\eta=\frac12 \ln \frac{t+x^1}{t-x^1}$ with $\tau_\eq$ being a reference proper-time; we choose it as the instant where the
hydrodynamical evolution starts. The transverse coordinate 
vector $\vec{x}_\perp = (x^2, x^3)$ is equal to the Cartesian one. In these coordinates, the non-zero Christoffel
symbols are
\begin{equation}
   \Gamma^0_{11} = \Gamma^0_{00} = \Gamma^1_{01} = \Gamma^1_{10}=1, \label{chr:01}
\end{equation}
and the metric tensor $g_{\mu\nu}$ is $diag(\tau_\eq^2 e^{2\Theta}, -\tau_\eq^2 e^{2\Theta}, -1, -1)$.
The constitutive equations can be used
together with the equation of state and the normalization condition of the four-velocity to obtain the
components of $T^{\mu\nu}$ as functions of four independent variables. We choose these variables to be $T^{\alpha 0}$, \ie
\begin{equation}
   T^{\mu\nu} = f^{\mu\nu}(T^{00}, T^{10}, T^{20}, T^{30}).\label{fmn:01}
\end{equation}
In principle, four arbitrary functions can be used to parametrize the components of the energy-momentum tensor.
Since we want to apply the iterative scheme to relativistic hydrodynamics, one could choose the energy density and
the spatial components of the four-velocity as well. There are two reasons why we not do so. First, setting up
the iterative scheme is most straightforward for this choice and second, we want to keep the discussion in this section
general, so the iterative scheme is not only applicable for ideal hydrodynamics, but also for other systems obeying
local energy-momentum conservation. However, there are some subtleties concerning energy-momentum tensors which contain
derivatives as in the case of viscous hydrodynamics. For such a situation \eqref{fmn:01} needs to be modified
(see appendix \ref{sec:appendix2}).
One important remark is that, when the setup is specified (\eg by choosing to investigate ideal hydrodynamics with a certain
equation of state), the
functional form of $f^{\mu\nu}$ is fixed and all its derivatives with respect to $T^{\alpha0}$ are given.
Equation \eqref{hyd:01} can be rewritten as
\begin{equation}
   \partial_0 T^{\mu 0} = -\partial_a f^{\mu a} - \Gamma^\mu_{\nu\rho}f^{\rho\nu} - \Gamma^\nu_{\sigma\nu}f^{\mu\sigma}. \label{hyd:03}
\end{equation}
Latin letters for indices refer to spatial coordinates ranging from 1 to 3.
Since $f^{\mu\nu}$ depends solely via its arguments $T^{\alpha 0}$ on the space-time coordinates,
the spatial derivative of $f^{\mu\nu}$ must be transformed into spatial derivatives of 
$T^{\alpha 0}$.
\par
The iterative scheme we want to describe is based on the Taylor expansion
\begin{equation}
   T^{\alpha 0}(\theta, \vec{\xi}) = \sum_{k=0}^{\infty} \frac{1}{k!}\, T^{\alpha0}_{(k)}(\vec{\xi})\,\theta^k,\label{ta0:01}
\end{equation}
where we have chosen without loss of generality coordinates for which the initial conditions are given at the
manifold $\theta=0$.
Analogue expansions hold for the Christoffel symbols and for $f^{\mu\nu}$:
\begin{equation}
   \Gamma^\alpha_{\beta\gamma} = \sum_{k=0}^{\infty} \frac{1}{k!}\, \Gamma^{\alpha}_{(k)\beta\gamma}(\vec{\xi})\,\theta^k,
   \hspace{1em}
   f^{\mu\nu} = \sum_{k=0}^{\infty} \frac{1}{k!}\, f^{\mu\nu}_{(k)}(\vec{\xi})\,\theta^k.\label{eq:fmn}
\end{equation}
Here $f^{\mu\nu}_{(k)}= \frac{\partial^k}{\partial {\theta}^k}f^{\mu\nu}(T^{\alpha 0}(\theta,\vec{\xi}))|_{\theta=0}$.
The time derivatives of $f^{\mu\nu}$
are then transformed according to the chain rule into time derivatives of $T^{\alpha 0}$ (see appendix \ref{sec:appendix1}).
Inserting these expansions into \eqref{hyd:03} yields
\begin{eqnarray}
   \sum_{k=0}^{\infty}&&\frac{1}{k!}T^{\mu0}_{(k+1)}{\theta}^k = 
   -\frac{\partial}{\partial \xi^a} \sum_{k=0}^{\infty}\frac{1}{k!} f^{\mu a}_{(k)}{\theta}^k\label{hyd:04}\\
   &&-\sum_{k=0}^{\infty}\sum_{l=0}^{k}\frac{1}{(k-l)!l!} \left(\Gamma^\mu_{(k-l)\nu\rho}f^{\rho\nu}_{(l)}
                                                +\Gamma^\nu_{(k-l)\sigma\nu}f^{\mu\sigma}_{(l)}\right){\theta}^k.\nonumber
\end{eqnarray}
It is important to note that
$f^{\mu\nu}_{(m)}$ contains only time derivatives of $T^{\alpha 0}$ up to order $m$. By comparing the coefficients, \eqref{hyd:04}
can be decomposed into a series of algebraic recurrence formulas relating the Taylor coefficients:
\begin{eqnarray}
   T^{\mu0}_{(k+1)} = 
   &-&\frac{\partial}{\partial \xi^a} f^{\mu a}_{(k)}\nonumber\\
   &-&\sum_{l=0}^{k}\binom{k}{l} \left(\Gamma^\mu_{(k-l)\nu\rho}f^{\rho\nu}_{(l)}
                                    +\Gamma^\nu_{(k-l)\sigma\nu}f^{\mu\sigma}_{(l)}\right)\pdown\label{hyd:05}
\end{eqnarray}
Accordingly, the $(k+1)$th Taylor coefficient $T^{\mu 0}_{(k+1)}$
is algebraically to be calculated by lower-order coefficients. This makes an iterative solution
possible.
The initial conditions
 determine $T^{\alpha 0}_{(0)}$. Using the equation with $k=0$ one can calculate 
$T^{\alpha 0}_{(1)}$ from that. Applying the equation with $k=1$ one can calculate $T^{\alpha 0}_{(2)}$ etc. 
In such a way, arbitrarily high Taylor coefficients are calculated iteratively.
\par
$T^{\alpha 0}_{(0)} = T^{\alpha 0}(\theta = 0, \vec{\xi})$ is supposed to be a smooth (differentiable) function.
The determination of high-order time derivatives is therefore via \eqref{hyd:05} reduced to calculating high-order spatial 
derivatives of $T^{\alpha 0}(\theta=0, \vec{\xi})$.
\par
For the $(\Theta, \eta, \vec{x_\perp})$ coordinate system, the Christoffel symbols \eqref{chr:01} are constants. Therefore
only the terms with $k=l$ contribute, and \eqref{hyd:05} reduces to
\begin{equation}
   T^{\mu0}_{(k+1)} = -\frac{\partial}{\partial \xi^a} f^{\mu a}_{(k)}
   -\left(\Gamma^\mu_{(0)\nu\rho}f^{\rho\nu}_{(k)}
                                    +\Gamma^\nu_{(0)\sigma\nu}f^{\mu\sigma}_{(k)}\right).\label{hyd:06}
\end{equation}

\section{Ideal fluid dynamics}
\label{sec:IdFl}
The scheme described in the preceding section is, in principle, applicable to any physical system obeying local 
energy-momentum conservation.
Therefore, it can be used to calculate solutions of the relativistic hydrodynamical equations.
For ideal fluid dynamics, where $\Pi^{\mu\nu}$ in \eqref{hyd:02} is neglected, 
the relation \eqref{fmn:01} becomes explicitly
\begin{equation}
   f^{ab}(T^{\alpha 0}) = f T^{a0} T^{b0} - p g^{ab}, \label{fmn:02}
\end{equation}
with
\begin{eqnarray}
   f  &=& \frac {g_{00}  \left( 5 A+2 g_{00} {T^{00}}^2 -T^{00} \sqrt {g_{00}}\sqrt {-5 A+4 g_{00} {T^{00}}^2 } \right) }
                 { \left( 3 g_{00} T^{00}+\sqrt {g_{00}}\sqrt {-5 A+4 g_{00} {T^{00}}^2 } \right)A},
          \nonumber\\[-5ex]\label{ideal:f}\\[2ex]
   A  &=& g_{ab}T^{a0}T^{b0},\\
   p &=&  {\frac { \left( g_{00} {T^{00}} ^2+A \right) g_{00}}
             {3 g_{00} T^{00}+\sqrt {g_{00}}\sqrt {-5 A+4 g_{00} {T^{00}} ^2}}}.\label{ideal:p}
\end{eqnarray}
By definition $f^{\mu 0} = f^{0\mu} = T^{\mu 0}$.
The four-velocity components $u^\mu$ are given by
\begin{eqnarray}
   u^0 &=&  {\frac {\sqrt {2 g_{00} {T^{00}}^2 - A + T^{00} \sqrt {g_{00}}\sqrt {-5  A+4 {T^{00}}^2g_{00}}}}
                   {2\sqrt {g_{00}}\sqrt { A+{T^{00}}^2g_{00}}}},\ \ \ \ \\
   u^a &=& hT^{a 0},\label{ideal:u_a}
\end{eqnarray}
with
\begin{eqnarray}
   h = \frac{\sqrt {5  A+2 g_{00} {T^{00}}^2-T^{00} \sqrt {g_{00}}\sqrt {-5  A+4 g_{00} {T^{00}}^2}}}
             {2\sqrt { A}\sqrt {g_{00} {T^{00}}^2+ A}}.\ \ \label{fmn:03}
\end{eqnarray}
The above equations hold for the conformal equation of state $e=3p$, but can be generalized to others in a straightforward manner.
They are obtained by considering the $(\mu\, 0)$ components of \eqref{hyd:02}. These four equations are solved for 
the pressure $p$ and the spatial components of the four velocity $u^a$ applying the equation of state and the normalization
of the four velocity. Therefore, one can express $p$ and $u^a$ as functions of the components $T^{\alpha 0}$ of the 
energy-momentum tensor (\cf \eqref{ideal:p} and \eqref{ideal:u_a}). The expressions $p(T^{\alpha 0})$, $u^a(T^{\alpha 0})$ 
are plugged into \eqref{hyd:02} from which the algebraic relations \eqref{ideal:f} can be read off by applying \eqref{fmn:01}.
We note that the equations \eqref{fmn:02} - \eqref{fmn:03} apply for any time-orthogonal coordinate system.
\section{Examples}
\label{sec:Examples}
The iterative scheme is implemented in MAPLE to calculate derivatives and consecutively the expansion coefficients.
\subsection{Longitudinal expansion}
Let us consider first a purely longitudinal expansion. Following \cite{Eskola:1997hz} we employ for the initial
conditions at \mbox{$\Theta_0 = 0$} a Gaussian energy distribution, $e(\Theta_0, \eta) 
= e_0 \exp\{-\frac{\eta^2}{2\sigma^2}\}$ with
$\sigma=3.8$ for LHC energies and a
synchronized flow $ y(\Theta_0, \eta)=\eta$. The latter one is important for the finding, already anticipated in
\cite{Eskola:1997hz}, that the energy density evolution can be approximated by
$e(\Theta, \eta)\approx e(\Theta_0,\eta) \exp\{-\frac{4}{3}\Theta\}$ and the flow velocity is only
modified on the 2\% level by a time dependent deviation from the Bjorken flow pattern characterized by $a(\Theta)$ according 
to $y = \eta(1+ a(\Theta))$. This statement is made more explicit below.
\par
\begin{figure}
   \centering
   \subfigure{\includegraphics[width=0.6\textwidth]{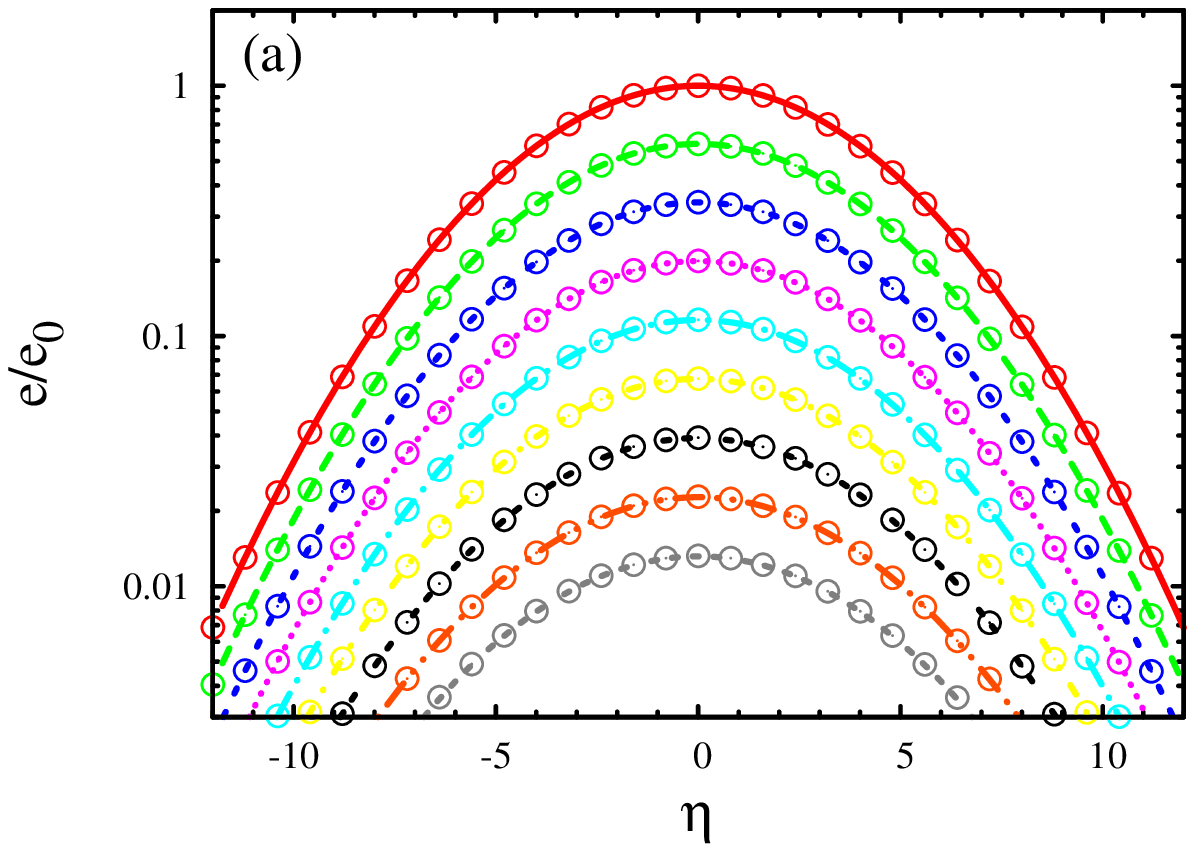}\label{fig:longED_log}}\\
   \subfigure{\includegraphics[width=0.6\textwidth]{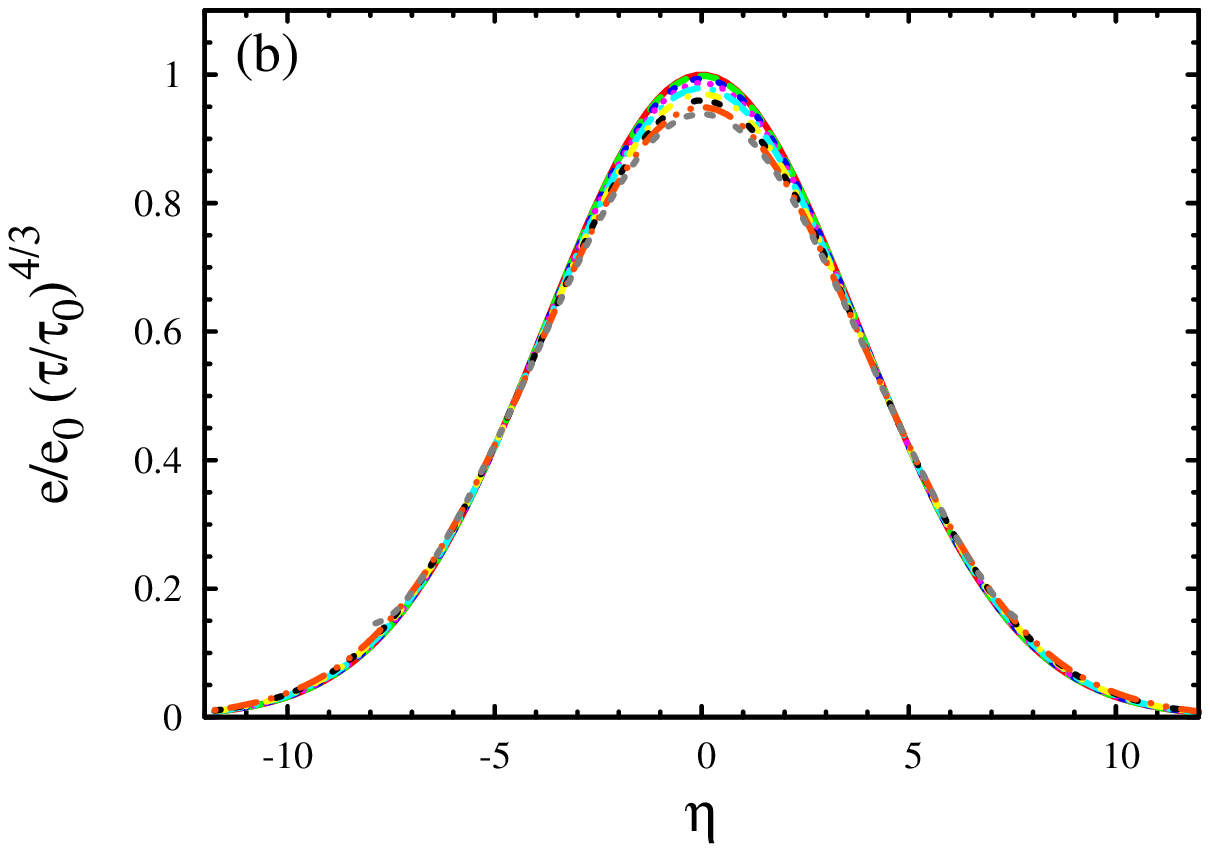}\label{fig:longED_BJ}}\\
   \subfigure{\includegraphics[width=0.6\textwidth]{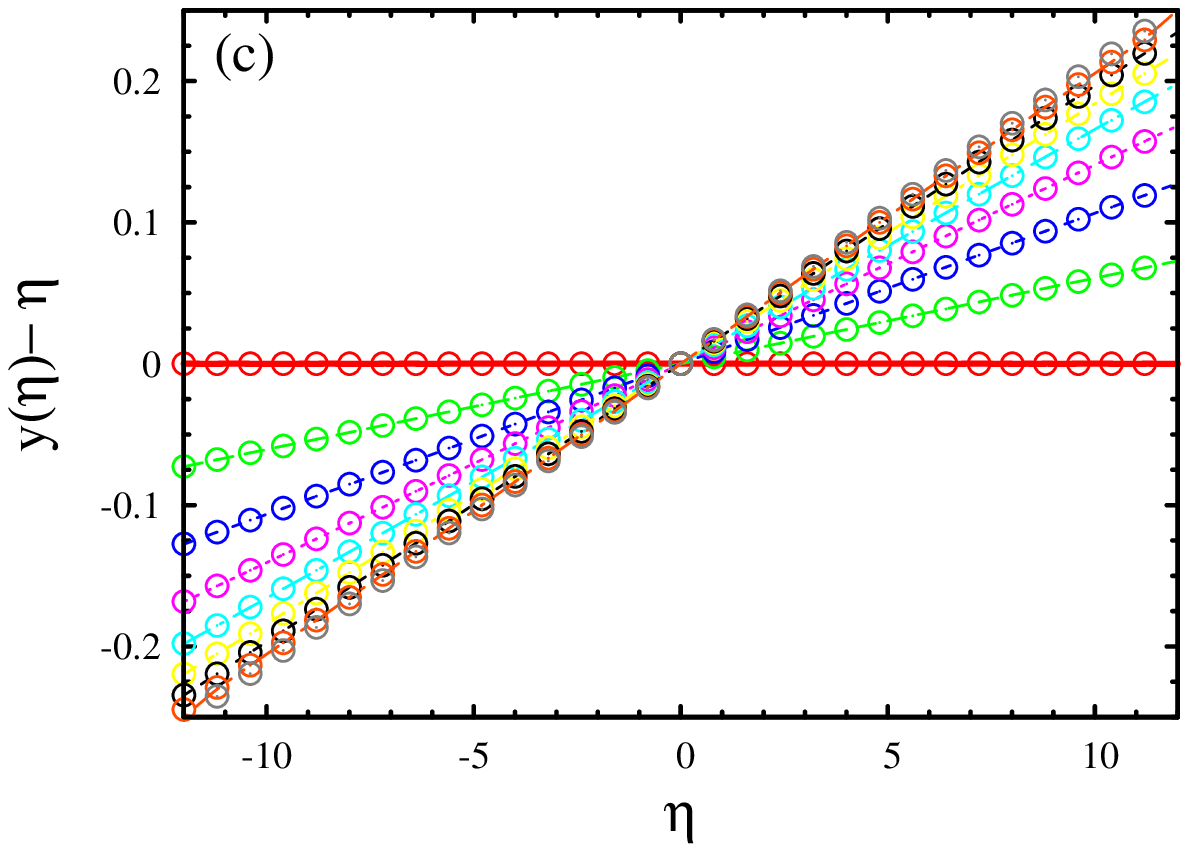}\label{fig:longV}}
   \caption{Energy density scaled either by $e_0=e(\Theta_0, \eta=0)$ (a) or by 
            $e_0(\tau_\eq/\tau)^{-4/3}= e_0 \exp\{-\frac{4}{3}\Theta\}$ (b) 
            and the difference of flow rapidity to space-time rapidity (c) for a 
            purely longitudinal expansion of an initially Gaussian energy distribution.
            The curves correspond to $\Theta=\ln(\tau/\tau_\eq) = 0.4, 0.8, 1.2, 1.6, 2.0, 2.4, 3.2$
            (in (a) from top to bottom and in (c) counterclockwise).
            The
            curves are the results of the iterative scheme up to order 48, and the circles depict a 
            numerical solution of the hydrodynamical equations by a differencing scheme.}
   \label{fig:long}
\end{figure}

Figure \ref{fig:long}
exhibits $e(\Theta,\eta)/e(\Theta_0,\eta=0)$ (a), $e(\Theta,\eta)/$ $e(\Theta_0, \eta=0)\exp\{\frac43 \Theta\}$
(b) and $y(\Theta,\eta)-\eta$ (c) as a function of $\eta$ for various time instants. The iterative scheme has been used up to order $N=48$. 
In figs. \ref{fig:long} (a) and (c), the iterative results (curves) are compared to a numerical solution of the hydrodynamical 
equations (\cf eqs. (10) and (11) in \cite{Eskola:1997hz}) obtained with a modified forward integration scheme (symbols).
One observes agreement of both solution methods up to $\Theta = 3.2$, \ie $\tau = \tau_\eq e^\Theta=24.5\,\tau_\eq$. At later
proper times $\tau$, the truncated iterative result shows some instability types, signaled by irregularities in $e(\eta)$ 
accompanied by negative energy densities at large values of $\eta$. However, these
can be separated from the physically interesting region. For larger values of $\Theta$, \eg $\Theta\approx 4.0$ 
(\ie $\tau\approx 55\,\tau_\eq$), 
these instabilities even influence the energy density at the center. 
But at this proper time, the medium has cooled down so far that hadronisation is almost complete.
In fig. \ref{fig:long} (c), the value of $a(\Theta)$ can be read off as the slope of the appropriate curve. Even for 
the largest depicted time $\Theta=3.2$ the slope $\Delta (y-\eta)/\Delta \eta$ being equal to $a(\Theta)$
is found to be small: $0.25/12 \approx 0.02$.
Since the slope of $y(\eta)$
increases monotonously with $\Theta$, the absolute value of $a(\Theta)$ for $0\leq \Theta \leq 3.2$ is always smaller than the 
above mentioned $2\%$. 
The conclusion for this simplified flow pattern is that the Bjorken type scaling of the energy density
$\propto \exp\{-\frac 43\Theta\}$ and synchronized flow $y-\eta\approx 0$ are excellent approximations
for mid-rapidity $|\eta| < 0.9$, accessible in the ALICE experiment. 
Analytical corrections to the Bjorken behavior that can be obtained utilizing the iterative scheme are discussed below.
\par 
Taking the initial temperature
as $T(\Theta_0, \eta\approx0)\approx 430\,$MeV corresponds to the energy density 
$e(\Theta_0, \eta\approx0)\sim 65\,\mathrm{GeV\,fm}^{-3}$ (obtained with the s95p-PCE equation of state in \cite{Shen:2010uy}; 
for lattice data see \eg \cite{Borsanyi:2010cj}).
Following the evolution until freeze-out at $T_{\fo}\sim 165\,$MeV, corresponding to the energy density of 
$e_{\fo}\sim 500\mathrm{\,MeV\,fm}^{-3}$, requires to go up to 
$\Theta = \ln(\tau/\tau_\eq )= \ln(e(\Theta_0)/e_{\fo})^{3/4}\approx 3.7$.
Figure \ref{fig:long} demonstrates that the iterative scheme, truncated at $N=48$ is stable in this time interval.
\par
\begin{figure}
   \centering
   \subfigure{\includegraphics[width=0.6\textwidth]{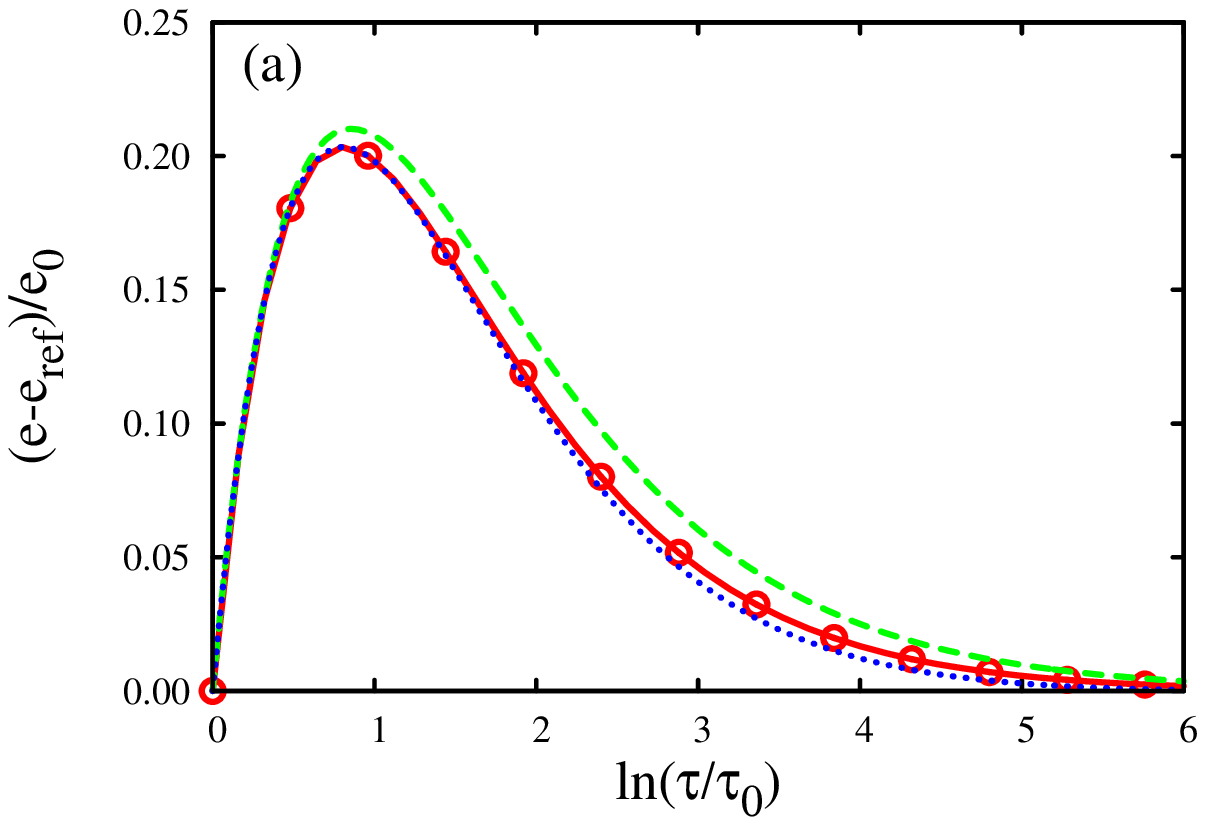}}\\
   \subfigure{\includegraphics[width=0.6\textwidth]{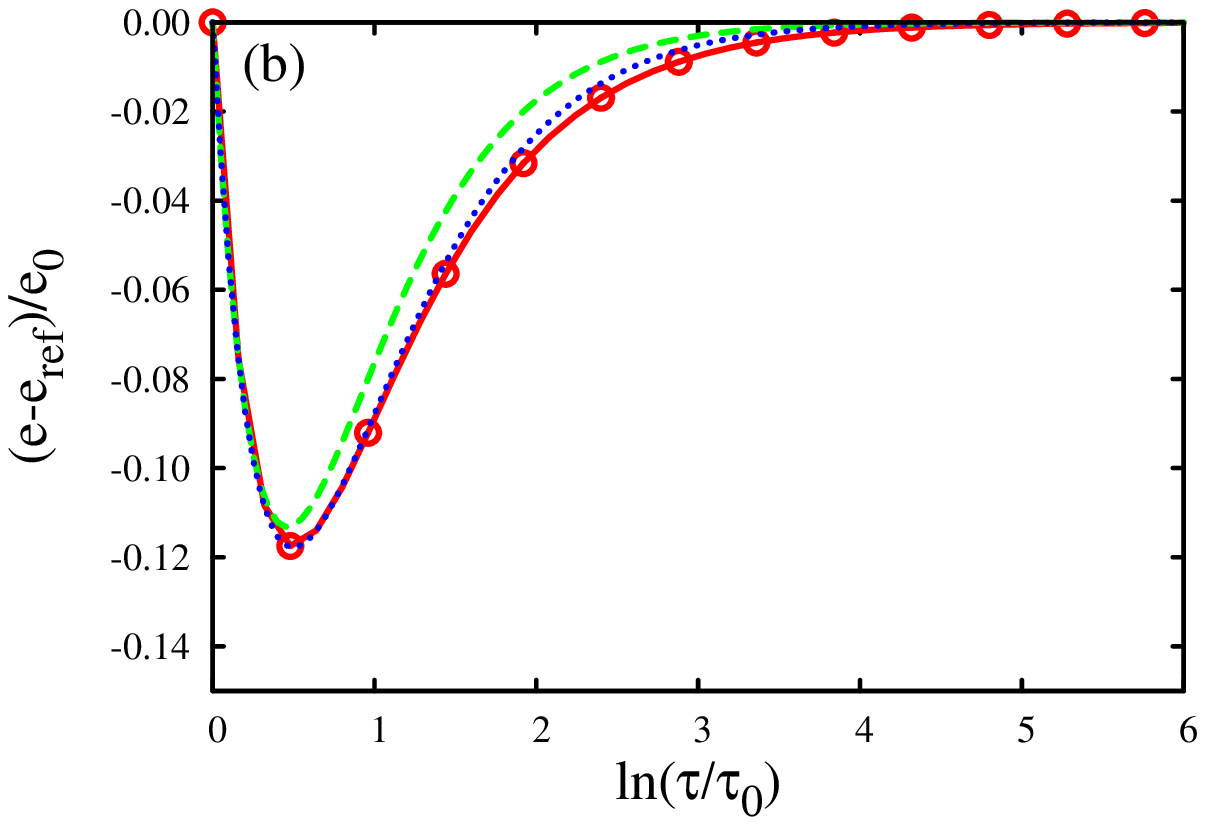}}
   \caption{Difference of the energy density $e$ of initially not synchronized flow $y(\Theta_0)=(1+a(\Theta_0))\eta$ 
            (with $a(\Theta_0)=-\frac12$ (a) and $a(\Theta_0)=\frac12$ (b)) 
            to the energy density $e_{\mathrm{ref}}$ of initially synchronized flow ($y(\Theta_0) = \eta$), evaluated at
            mid-rapidity as functions of the time parameter $\Theta=\ln(\tau/\tau_\eq)$. In
            both cases, the initial energy density is Gaussian 
            $e(\Theta_0) = e_{\mathrm{ref}}(\Theta_0)=e_0\exp\{-\eta^2/(2 \sigma^2)\}$ with $\sigma= 3.8$.
            The green dashed curves are based on \eqref{dev:02} evaluated at leading order, and the blue dotted curves are the 
            corrections
            evaluated at next-to-leading order. The curves are compared to the numerical solution (red solid curves with symbols).
            }
   \label{fig:approx}
\end{figure}

To expose the importance of the initially synchronized Bjorken type flow pattern we chose 
$y(\Theta_0, \eta) = (1+a(\Theta_0))\eta$.
For values of $a(\Theta_0)$ not to far away from the Bjorken flow (\ie $a(\Theta_0)=-1\dots0.5$),
the deviation of the energy density from the reference system (a system with initially synchronized flow, \ie $a(\Theta_0)=0$)
can be parametrized by the series
\begin{equation}
   e-e_{\mathrm{ref}} \approx \sum_n d_n \Theta^n e^{-b_n \Theta}. \label{dev:01}
\end{equation}
Here, $e$ denotes the energy density of the system with initially not synchronized flow, and $e_{\mathrm{ref}}$
is the energy density of a reference system with the same initial energy distribution, but Bjorken type initial flow.
The parameters in this expansion can be directly linked to the Taylor coefficients of the energy density, by
differentiating \eqref{dev:01} and evaluating the result at $\Theta_0$. Apart from constant factors, the left hand side
is the difference of the Taylor coefficients of $e$ and $e_{\mathrm{ref}}$, which are calculable within the iterative
scheme. The right hand side gives algebraic expressions for the parameters $d_n$ and $b_n$.
This method is quite powerful and can be used to transform the iterative solution given in terms of Taylor coefficients
into an equivalent solution in terms of a different set of basis functions than powers of $\Theta$ 
(\eg those given in \eqref{dev:01}).
For a system with non synchronized flow and initially Gaussian energy distribution, 
it turns out that the first non-vanishing terms already give excellent agreement:
\begin{equation}
   e - e_{\mathrm{ref}} \approx d_1 \Theta e^{-b_1 \Theta} + d_3 \Theta^3 e^{-b_3 \Theta}.\label{dev:02}
\end{equation}
For general values of $\eta$, the expressions are quite cumbersome; at mid-rapidity they reduce to
\begin{eqnarray}
   d_1 &=& -\frac{4}{3}a,\\
   b_1 &=& \frac{5}{3} + a,\\
   d_3 &=& -\frac{2a(-15+14 a \sigma^2+17 \sigma^2 a^2+\sigma^2)}{81 \sigma^2 },\\
   b_3 &=& \frac{298 \sigma^2 a^2-120 a-165+166 a \sigma^2+10 \sigma^2+126 \sigma^2 a^3}
                {6(-15+14 a \sigma^2+17 \sigma^2 a^2+\sigma^2)},\nonumber\\[-2ex]
\end{eqnarray}
with $a\equiv a(\Theta_0)$ for ease of the notation.
Figure \ref{fig:approx} illustrates this approximation. One observes that at mid-rapidity a larger value of $a(\Theta_0)$ 
(\ie a larger initial flow)
leads to negative corrections with respect to an initially synchronized flow, meaning a more rapid decrease of central
energy density.
At large times one sees the difference approaching zero, which is a consequence of the energy density approaching
zero. But if the initial flow is realistic (\ie close to synchronized flow, $a(\Theta_0)\approx 0$), the difference decreases even
faster (\ie with a larger decay rate in the exponential) than the energy density itself (which decreases approximately 
like the Bjorken energy density $e_\BJ = e_0 e^{-4\Theta/3}$). That means that the solution
stabilizes around the solution with initially synchronized flow, and small perturbations have little effect on the 
long-time behavior. Since for realistic initial conditions ($a(\Theta_0)$ close to zero and $\sigma\approx 4$ for LHC) already the 
leading-order correction is very good for all times until freeze-out, one can calculate the time and the value of the largest
deviation from the reference solution. The maximum of $e - e_{\mathrm{ref}} \approx d_1 \Theta e^{-b_1\Theta}$ is at 
$\Theta=1/b_1$, where the deviation has the value of $ (d_1/b_1)e^{-1}$.
\par
A similar analysis can be done to study the dependence of the energy density on the width of the initial energy distribution.
It is found that the deviation of the energy density $e$ obtained with a Gaussian shaped (standard deviation: $\sigma$) 
initial energy density from the
Bjorken energy density $e_\BJ$ can also be described with the series \eqref{dev:01}. In this case, the leading terms are
\begin{equation}
   \frac{e(\Theta, \eta)}{e(\Theta_0, \eta)}-e_{\BJ} \approx d_2 \Theta^2 e^{-b_2 \Theta} 
                                                          + d_4 \Theta^4 e^{-b_4 \Theta}.\label{dev:03}
\end{equation}
\begin{figure}
   \centering
   \includegraphics[width=0.6\textwidth]{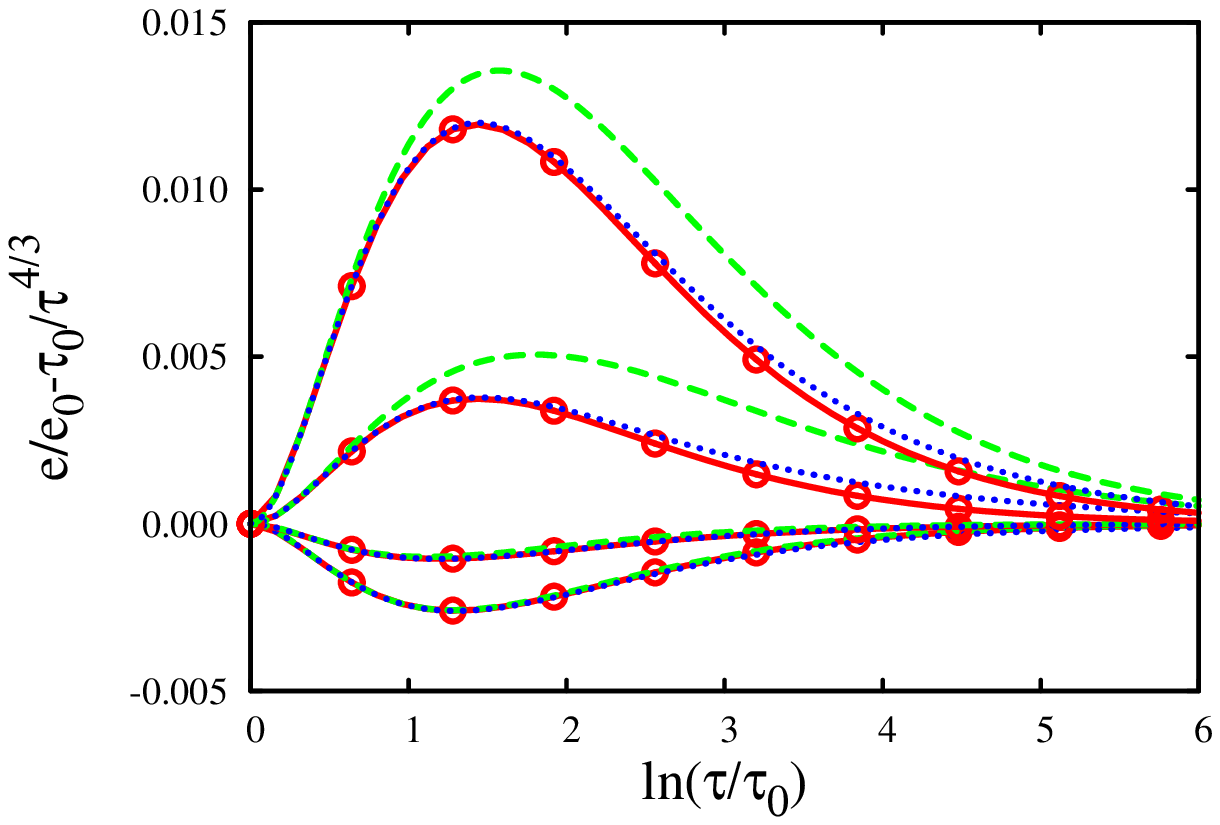}
   \caption{Difference of the energy density $e$ (normalized to its initial values $e_0=e_0(\eta)$)  of initially Gaussian 
            shaped energy density to the Bjorken energy density as a function of $\Theta=\ln(\tau/\tau_0)$.
            The initial flow is chosen to be synchronized $y = \eta$. The different curves correspond to
            different longitudinal positions (from bottom to top: $\eta = 0, \sigma, 2\sigma, 3\sigma$).
            The green dashed curves are based on \eqref{dev:03} evaluated at leading order, and the blue dotted curves are 
            the corrections
            evaluated at next-to-leading order. The red solid curves with symbols depict the numerical solution
            obtained with a forward integration scheme.
            }
   \label{fig:approx2}
\end{figure}
The normalisation to the initial energy density is necessary to disentangle effects originating from gradients
from the effect of different initial values at different spatial coordinates.
Since the expressions for $d_4$ and $b_4$ at a generic $\eta$ are quite extensive
this coefficients are only given at $\eta=0$:
\begin{eqnarray}
   d_2 &=& \frac{-2\sigma^2+\eta^2}{12\sigma^4},\\
   b_2 &=& \frac{4(-7 \sigma^2+3 \eta^2)}{9(-2 \sigma^2+\eta^2)},\\
   d_4(\eta=0) &=& -\frac{4 \sigma^2-45}{1944\sigma^4},\\
   b_4(\eta=0) &=& \frac{2(-5265+428 \sigma^2)}{135(4 \sigma^2-45)}.
\end{eqnarray}
For LHC conditions the leading order term is much more important than the subleading order term 
(except at $\eta \approx \sqrt{2}\sigma$, where it vanishes), \ie the leading term gives already good
agreement. In this case, the position and the value of the largest deviation $e/e(\Theta_0) - e_{\mathrm{ref}}$
can be calculated analytically, and one
finds that the maximum value of $(4 d_2/b_2^2)e^{-2}$ is taken at $\Theta = 2/b_2$. In Fig. \ref{fig:approx2},
the approximations are plotted at several values of $\eta$. One notes that the next-to-leading order approximation
gives excellent agreement with the numerical data and the leading-order approximation is also good.
\subsection{Transverse expansion: axisymmetric flow}
\begin{figure}
   \centering
   \includegraphics[width=0.6\textwidth]{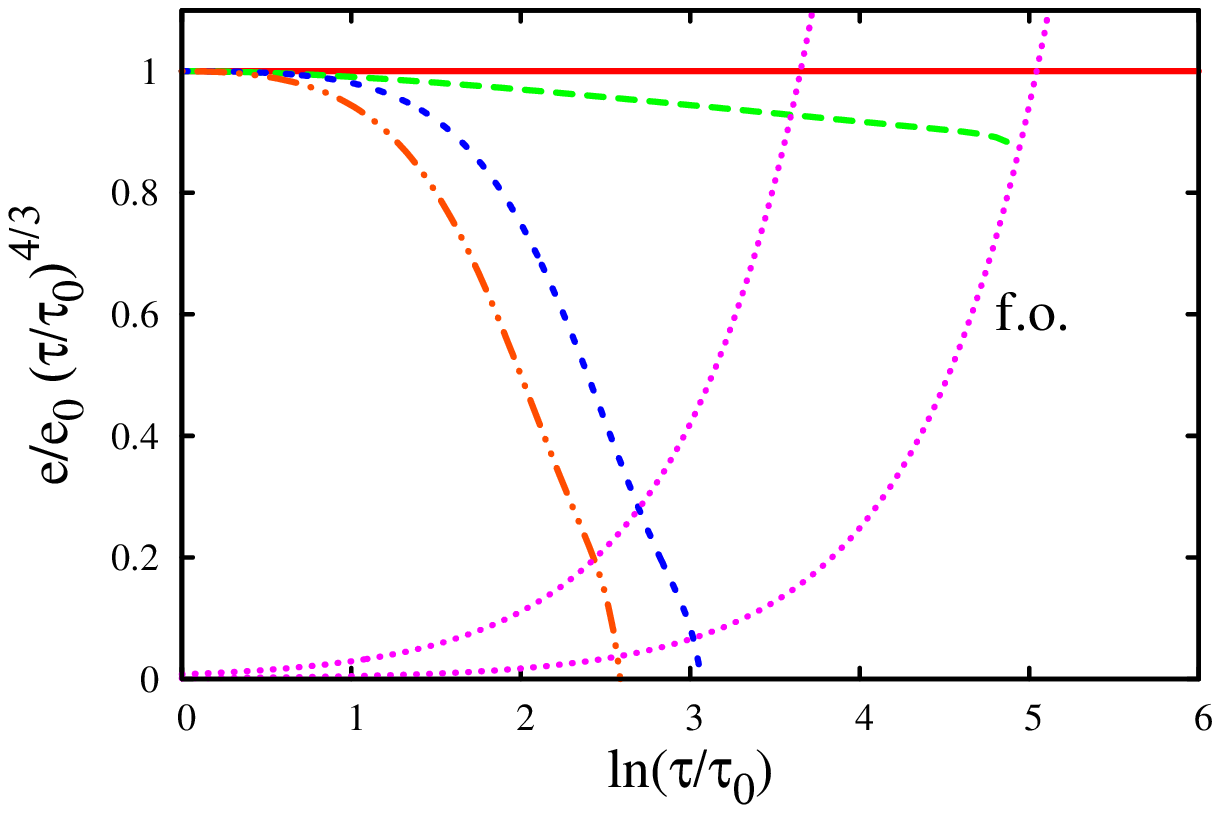}
   \caption{Scaled energy densities at the center $\vec{x_\perp}=0$ at mid-rapidity for the Bjorken solution (solid red curve)
            and the longitudinal (green dashed curve) as well as axisymmetric
            transversal (blue dotted curve) and triaxial (orange dash-dotted curve) expansion as a function of time 
            $\Theta= \ln(\tau/\tau_\eq)$. 
            The truncation orders are $N= 48,48$ and $24$ for the longitudinal, transversal and triaxial expansion respectively.
            The freeze-out curves correspond to freeze-out temperatures $T_{\fo}=165\,$MeV (left purple dotted line) 
            and $T_{\fo}=130\,$MeV (right purple dotted line). The central
            energy density at initial time is chosen to be $e_0=65\,\mathrm{GeV fm^{-3}}$, corresponding to LHC energies.}
   \label{fig:ed_center}
\end{figure}

Equipped with the finding of the previous subsection
 we now turn to the axisymmetrical transverse expansion. In doing so, we assume for the longitudinal flow 
$y(\Theta, \eta)=\eta$ consistent with $e(\Theta,\eta, r)=e(\Theta,r)$ \ie longitudinal scaling symmetry,
which implies an infinitely large extension in longitudinal direction. 
The initial transverse energy density distribution is 
$e(\Theta_0, r) = e_0 \exp\{-r^2/(2\kappa^2)\}$ with the transverse coordinate $r=\sqrt{\vec{x_\perp}^2}$ and 
$\kappa$ being the standard deviation of the distribution.
\begin{figure}
   \centering
   \subfigure{\includegraphics[width=0.6\textwidth]{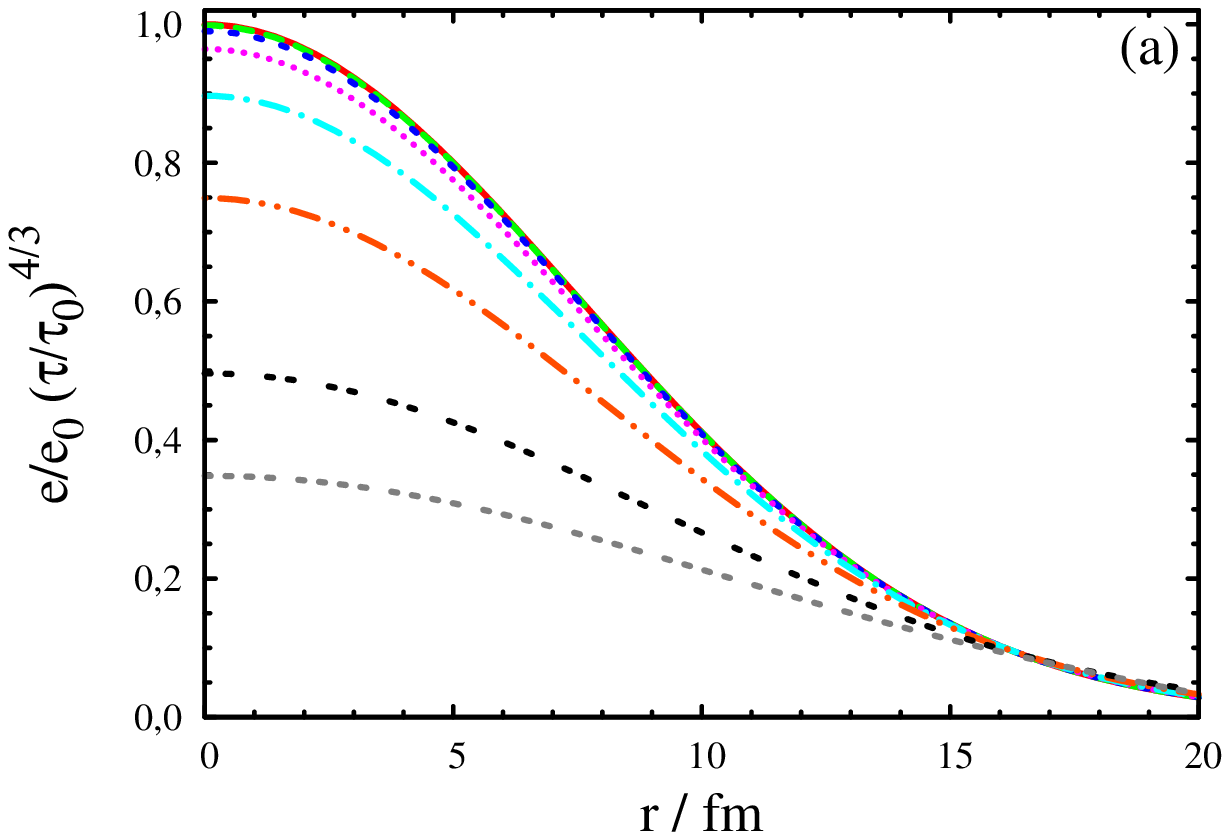}\label{fig:transED_BJ}}\\
   \subfigure{\includegraphics[width=0.6\textwidth]{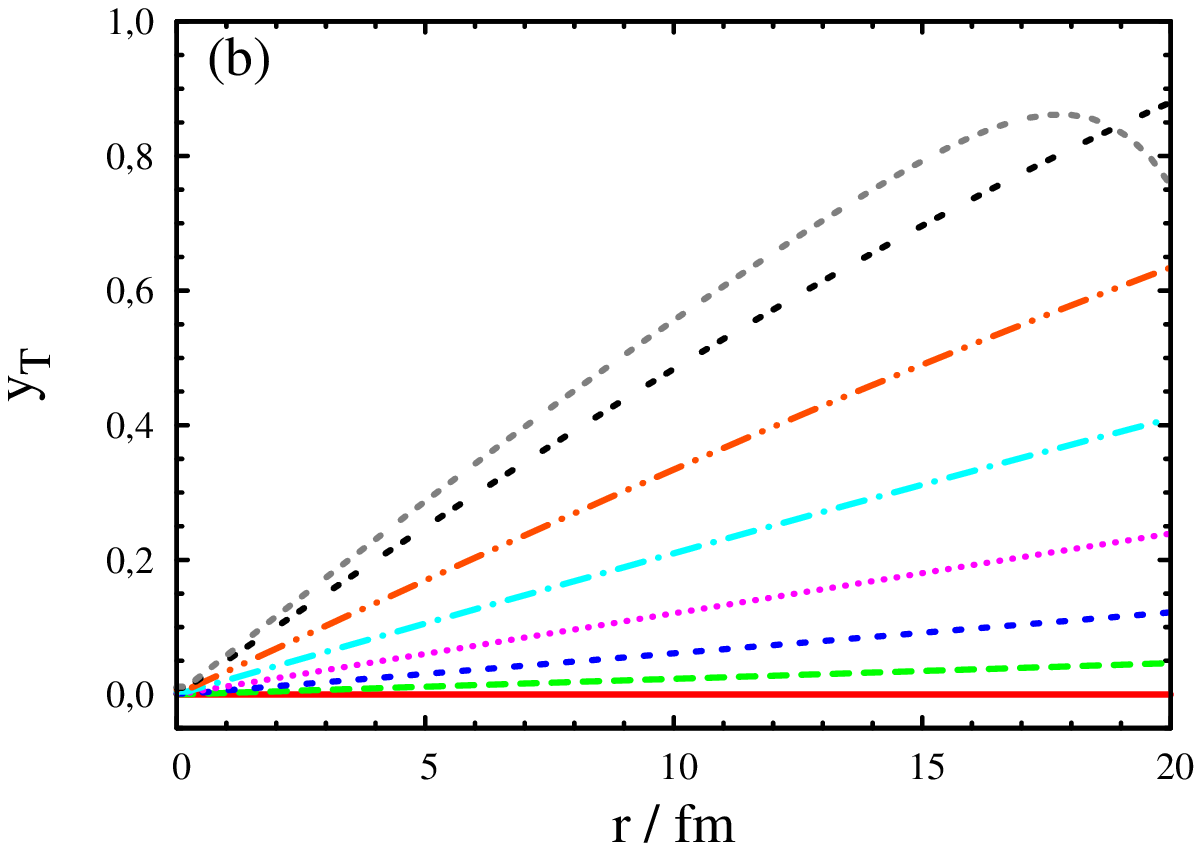}\label{fig:transV}}
   \caption{Scaled energy density (a) and transversal flow rapidity $y_T= \mathrm{arctanh}(v_T)$ (b), with $v_T$
            being the velocity in the transversal direction for a 
            transversal expansion of an initially transversal Gaussian energy distribution superimposed on the scaling 
            invariant longitudinal flow.
            The times shown are $\Theta=\ln(\tau/\tau_\eq) = 0.4, 0.8, 1.2, 1.6, 2.0, 2.4, 2.6$ (from top to bottom in
            (a) and from bottom to top in (b)). The
            curves are the results of an iterative calculation up to order $N=48$.}
   \label{fig:trans}
\end{figure}
This transverse distribution is certainly to smooth and therefore underestimates the transverse gradients. In 
this respect it may give some lower limit on the transverse flow which is initially zero.
\mbox{Figure \ref{fig:trans}} shows the evolution of the scaled energy density and the transversal flow rapidity for a 
distribution
with initial width $\kappa=7.5\,$fm corresponding roughly to the radius of lead nuclei used for the heavy-ion experiments
at the LHC.
The transversal expansion has a much larger effect on the dynamics
than it was the case for the longitudinal expansion, since it is not dominated by the initial flow. 
For instance, at mid-rapidity, the scaled energy density drops by $\approx 5\,\%$ for $\Theta=\ln(\tau/\tau_\eq)=2.5$ 
in the case of purely 
longitudinal expansion(\cf green dashed curve in Fig. \ref{fig:ed_center}), while for the transverse expansion, 
the scaled energy density is reduced by $58\,\%$ at the same time
(\cf blue dotted curve in Fig. \ref{fig:ed_center}). 
\par 
The backbending of the rapidity curve in Fig. \ref{fig:trans} (b) for $\Theta=2.6$ may be attributed to instabilities of the 
truncated iterative scheme for larger times.
However, the energy density at the corresponding space-time region is already below the freeze-out energy density. 
\subsection{Transverse expansion: asymmetric flow}
\label{sec:trans_as}
\begin{figure}
   \centering
   \includegraphics[width=0.6\textwidth]{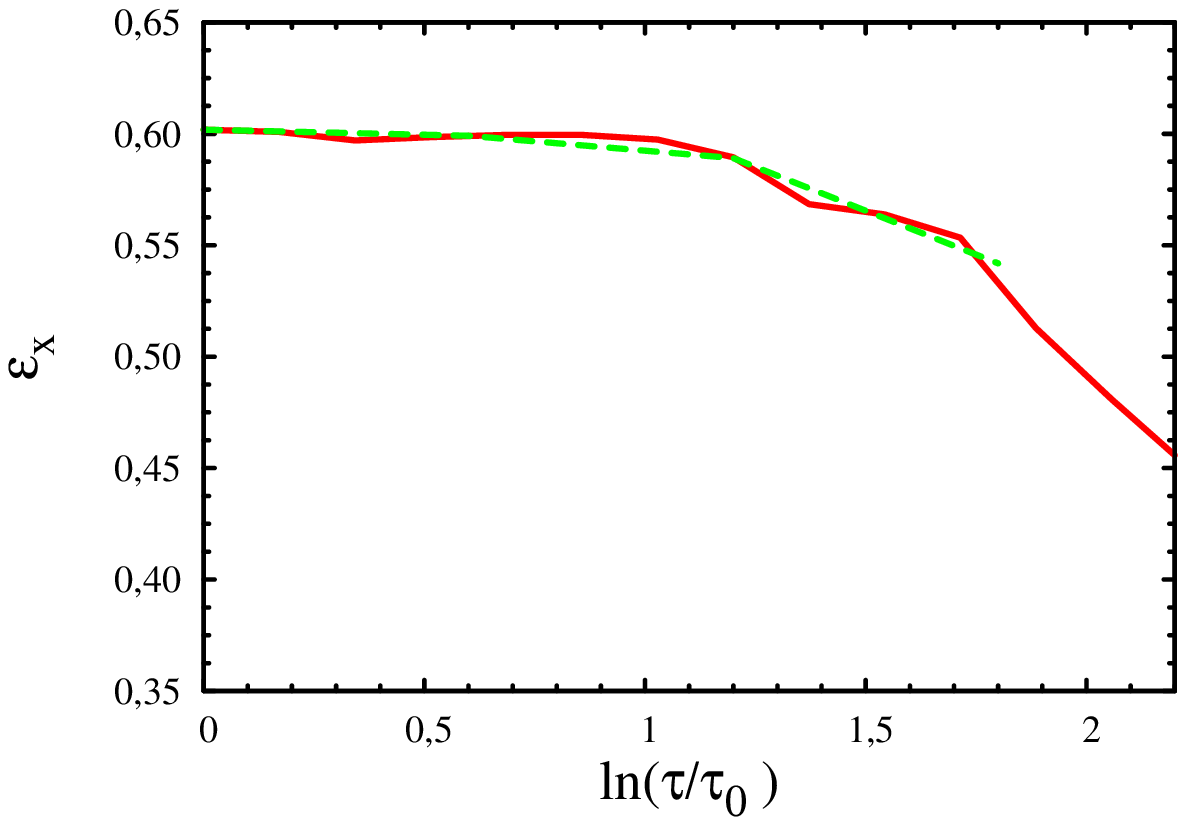}
   \caption{Spatial eccentricity at mid-rapidity of the energy density distribution with infinite size in longitudinal 
            direction from a calculation up to order $N=43$ (red solid line) and the same quantity 
            for an additionally longitudinal Gaussian shaped energy density calculated up to order $N=24$ 
            (green dashed line). The latter energy density suffers earlier from instabilities and is therefore
            only depicted until $\Theta=\ln(\tau/\tau_\eq)=1.8$.}
   \label{fig:Ex}
\end{figure}

Transport properties of the sQGP can be extracted from the evolution of initial asymmetries.
The most intensively studied quantity that encodes such information is the elliptic flow $v_2$, which is used to
measure, \textit{e.g.}, the shear viscosity of the sQGP. Therefore it is interesting to check whether the iterative scheme
can cope with asymmetric initial conditions and if the expected change of eccentricity can be observed
in the results. For this purpose, initial conditions 
$e(\Theta_0, x_2, x_3)= e_0 \exp\{-x_2^2/(2\sigma_2^2)-x_3^2/(2\sigma_3^2)\}$ 
with different widths $\sigma_2 = 7.5\,$fm and $\sigma_3 = 3.75\,$fm, respectively, along the two transversal axes were
used. 
The choice of $\sigma_2 : \sigma_3 = 2 : 1$ corresponds to an impact parameter of $b=6/5 R\approx 9\,$fm for lead nuclei
(in the hard sphere approximation).
% matching to a centrality of about 35\%.
As in the previous example the initial flow is purely longitudinal with 
$y(\Theta_0)=\eta$, and the longitudinal scaling symmetry is kept during the expansion implying independence of the
energy distribution on the longitudinal coordinate $\eta$.
In Fig. \ref{fig:Ex}, the resulting eccentricity
\begin{equation}
   \epsilon \equiv \frac{\int dx_2 dx_3 e(x_2, x_3)(x_2^2-x_3^2)}{\int dx_2 dx_3 e(x_2, x_3)(x_2^2 + x_3^2)}\label{ecc:01},
\end{equation}
with $e(x_2, x_3)$ from a calculation up to order $N=43$ is plotted (solid red curve).
\subsection{Triaxial expansion}
Even triaxial expansion can be treated within the iterative scheme. 
Since quantities like the elliptic flow are measured at mid-rapidity
and since in the vicinity of $\eta=0$ the energy density does not change much in the longitudinal direction, no 
significant change is expected for the eccentricities at $\eta=0$. The dashed green curve in Fig. \ref{fig:Ex} is obtained from
a calculation with triaxial initial energy density distribution 
$e(\Theta_0, \eta, x_2, x_3)= e_0 \exp\{-\eta^2/(2\sigma^2)-x_2^2/(2\sigma_2^2)-x_3^2/(2\sigma_3^2)\}$
with $\sigma= 3.8$, $\sigma_2 = 7.5\,$fm and $\sigma_3 = 3.75\,$fm and initially synchronized Bjorken flow, \ie 
$y(\Theta_0, \eta)=\eta$ and $\vec{v_\perp}=0$. It follows nicely the curve obtained for 
longitudinal scaling expansion, which shows that the spatial eccentricity and thereby the elliptic flow at mid-rapidity
is not sensitive to the details of the longitudinal flow. However, at $\Theta\gtrsim 1.9$ instabilities due to truncation 
make a comparison difficult. Curing this either by calculating at higher 
truncation order, optimizing the code or switching to an expansion similar to \eqref{dev:01} remains to be done and is left
to further studies.
\section{Limitations}
\label{sec:Limits}
If it would be possible to calculate the Taylor coefficients up to infinite order, the solution would be exact, at least
for analytical initial conditions and a smooth equation of state. However, due to 
limited computation power the iteration must be truncated at finite order $N$. As a result, only a truncated series is obtained and
errors occur that are at least one order higher than the truncation order. This is not necessarily a problem, since in case of
heavy-ion collisions one is only interested in a result covering a limited time span, for which such corrections may be small.
\par
\begin{figure}
   \centering
   \includegraphics[width=0.6\textwidth]{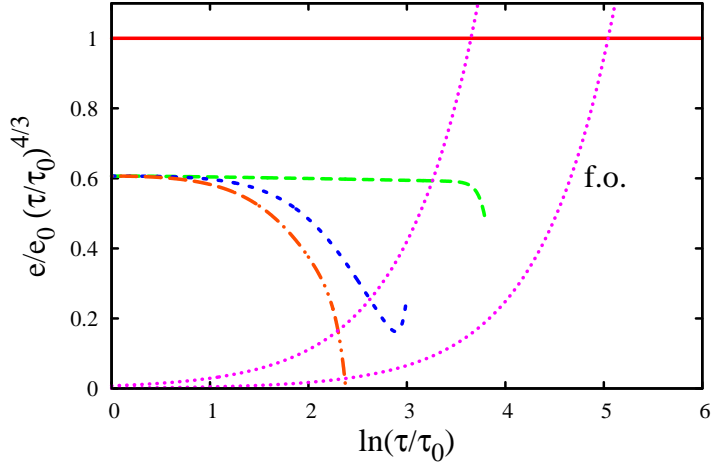}
   \caption{Energy density at the edges of the initial distribution. The freeze-out curves and initial conditions 
            as well as the truncation orders are the
            same as in Fig. \ref{fig:ed_center}. The longitudinal solution was evaluated at $\eta=\sigma=3.8$ and the 
            axisymmetric
            transversal solution at $r=\kappa=7.5$\,fm. Since the most interesting part of the triaxial expansion is the 
            behavior at mid-rapidity it was evaluated at $\eta=0$ and $x_2=0$ and $x_3 = 3.75$\,fm (\ie the steepest point
            of the initial energy distribution in the transversal plane at mid-rapidity).}
   \label{fig:ed_edge}
\end{figure}

In Fig. \ref{fig:ed_center}, the energy densities at the center of the fireball for the examples discussed above are
plotted together with the freeze-out region for typical LHC initial conditions.
One observes two facts. First, the central energy density of the purely longitudinal expanding system does not differ
much from the Bjorken solution. This is seen already in Fig. \ref{fig:long} (b). Second, the inclusion of transversal 
expansion leads to a much faster cooling of the medium since it expands in more directions.
Figure \ref{fig:ed_center} shows that the iterative scheme gives reasonable results for the energy density at the center 
of the fireball.
However, more problematic than the center are the regions where the gradients are large. To make this statement explicit,
in Fig. \ref{fig:ed_edge},
for each of the examples discussed above, the energy density at the edges is plotted. As representative for the edge, (one of) 
the point(s) is chosen, where the initial conditions have the largest gradients. One notes that the truncated iterative 
solutions get unstable quite close to the freeze-out region. This is an artefact of the truncated
iterative scheme, which demands for calculation with a low freeze-out temperature of $\sim 130\,$MeV a higher truncation order.
\par 
It is interesting to find out which properties of the initial condition improve the iterative result and which have a 
negative effect on it. Of course, the result gets better if one can calculate very high orders. To achieve
this one needs to use initial conditions that can be differentiated most easily. For this reason, Gaussian energy distributions
were used. An example for initial conditions that are not well suited for the iterative method are the ones deduced by Gubser 
\cite{Gubser:2010ze}. 
To understand the computational limits, one can estimate the number $n_k$ of terms involved in the $k$th Taylor coefficient 
if the initial conditions are of Gaussian type. One observes that the Taylor coefficients have the structure of an exponential
multiplied with a polynomial containing basically all spatial variables and all parameters in all powers up to $k$. This leads to 
the estimate, that $n_k$ grows like $k^{m_1+m_2}$ with $m_1$ being the number of spatial variables and $m_2$ being the number of 
parameters. 
The total number of terms involved in the Taylor series truncated at order $N$ is therefore 
$\sim \sum_{k=1}^N k^{m_1+m_2} \sim N^{m_1+m_2+1}$.
\par
\begin{figure}
   \centering
   \includegraphics[width=0.6\textwidth]{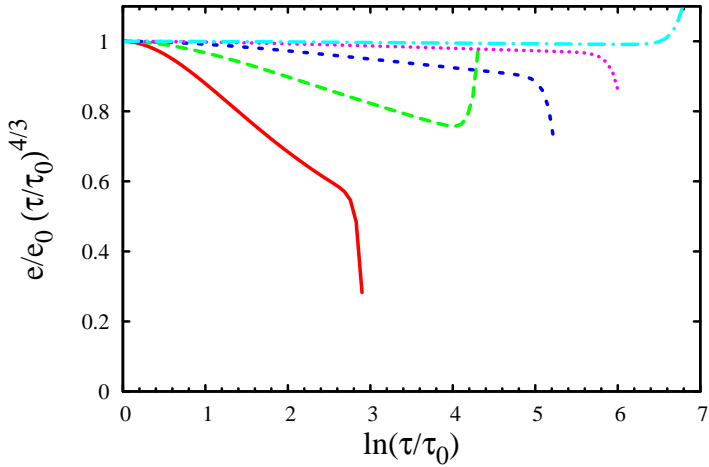}
   \caption{Energy density at the center at mid-rapidity for different widths of the initial energy distribution.
            The kink in the curves denotes the time, when the truncated iterative solution gets unstable.
            The initial energy distribution is $e(\Theta_0)= \exp\{-\eta^2/(2\sigma^2)\}$ 
            with $\sigma=1, 2, 4, 8, 16$ (from bottom to top).}
   \label{fig:ed_kink}
\end{figure}

  
A typical operation in the iterative evaluation is the multiplication of two of such series. For such a multiplication every
term in one series needs to be multiplied with one in the other, leading to $N^{2(m_1+m_2+1)}$ terms before simplification. 
For the transversal asymmetric
system discussed in section \ref{sec:trans_as} this results in a number of terms which is in the order of 
$43^{2\cdot 3}\approx6\times10^9$. Therefore, considering
additional dimensions or letting parameters general leads to a rapidly increasing demand of computation power and
reduces the calculable order dramatically.
A second way to obtain results that are close to the complete solution is looking for initial conditions
that lead to a quickly converging Taylor series. In this case, less orders need to be calculated, which obviously helps the
computation. 
Testing different initial shapes of the energy profile we realized that the convergence is better if the initial conditions
are more flat. Steep initial profiles lead to rapid oscillations in iterative results which can only be pushed to larger
times by computing considerably higher orders.  
A typical example for this behavior is shown in fig. \ref{fig:ed_kink}.
The kink that denotes the point, where instabilities of the truncated scheme modify the energy density at the central point,
occurs the later the broader the initial Gaussian is. 
\section{Conclusions}
\label{sec:Conclusion}
In this paper we present a scheme that can be used to construct solutions of the equations of relativistic 
fluid dynamics for differentiable initial conditions. The scheme is based on a formal Taylor expansion of 
the components of the energy-momentum tensor which is used to reformulate the time evolution equations as
an infinite set of algebraic relations between the Taylor coefficients that can be solved iteratively.
In the context of AdS/CFT similar methods are used to obtain solutions of the Einstein equations close to
the boundary of (asymptotic) anti-de Sitter space \cite{deHaro:2000xn,Janik:2010we}, but to our best knowledge
it was not applied to hydrodynamics or heavy-ion collisions in the literature so far. 
The scheme is utilized to address various cases of physical interest. It was shown that for the case of purely longitudinal
expansion the results obtained with the iterative scheme are in excellent agreement with the numerical solution
obtained with a standard numerical integration routine.
One can parametrize the difference between a solution and a reference solution in terms of a
special set of basis functions $\{d_n\Theta^ne^{-b_n\Theta}\}$. The parameters are related to Taylor
coefficients obtained within the iterative scheme and it was found that already the leading order terms give
very good agreement with the numerical solution obtained with a standard integration scheme.
\par
For transversal expansion it is possible to find solutions for the axial symmetric as well as for the axial asymmetric case.
In the latter case, the change of the shape of the energy density due to elliptic flow has been considered.
Even triaxial expansion needed for more realistic simulations is accessible within the iterative scheme.
At mid-rapidity the longitudinal shape of the medium has almost no influence on the transversal eccentricity, but
at finite $\eta$ it is expected to have an effect. 
\par
Concluding we can state that the iterative scheme is a promising tool
to investigate the properties of the sQGP produced in heavy-ion collision at the LHC via signatures built up 
during the hydrodynamic part of its evolution. With the computational power available for a personal computer at present,
its application to fluctuating initial conditions or other systems with initially large spatial gradients is somewhat limited.
Nevertheless there are several applications thinkable, \eg the calculation of the background field in which a jet is moving 
(\cf the use of hydrodynamics in \cite{Bass:2008rv}) or the emission of real and virtual photons as in \cite{Rapp:2011is}. 
Even an application as a benchmark is possible, since the result is exact up to the 
truncation order and should therefore match the results of other hydro-codes.  
\appendix
\section{Energy-momentum tensor with derivatives}
\label{sec:appendix2}
Several systems are characterized by an energy-momentum tensor containing derivatives of the basic physical
degrees of freedom. One important example is viscous relativistic hydrodynamics, \eg formulated in the Landau-Lifschitz frame.
For such an energy-momentum tensor it is not
possible to express all components as functions of $T^{\alpha 0}$ only. One possible way to cope with that problem is to
allow also derivatives of $T^{\alpha 0}$ as arguments, \ie
\begin{equation}
   T^{\mu\nu} = f^{\mu\nu}\left(T^{\alpha 0}, {T^{\alpha 0}}_{;\beta}, {T^{\alpha 0}}_{;\beta;\gamma}, \dots \right).
                     \label{fmn:04}
\end{equation}
with $;$ denoting the covariant derivative.
From \eqref{fmn:04} one can follow the steps outlined in section \ref{sec:Bas}. Another possible way would be to construct 
a similar iterative scheme directly for the basic physical degrees of freedom, \ie in the case of viscous relativistic
hydrodynamics for the energy density and the spatial components of the four-velocity.
\section{Faà di Brunos formula}
\label{sec:appendix1}
The Taylor expansion \eqref{eq:fmn} contains time derivatives of $f^{\mu\nu}$, evaluated at the initial time.
Since $f^{\mu\nu}$ does not depend explicitly on the coordinates the only time dependence originates from the time dependence
of its arguments $T^{\alpha 0}$.
With Faà di Brunos formula \cite{Knuth:73} it can be decomposed into derivatives of $f^{\mu\nu}$ 
with respect to $T^{\alpha 0}$ and derivatives of $T^{\alpha 0}$ with respect to $\theta$:
\begin{eqnarray}
   &&D^n(f\circ g) =\nonumber\\
   &&\ \sum \quad \frac{n!}{k_1!\cdot \cdots \cdot k_n!}(D^{k_1+\cdots+k_n}f)\circ g 
                   \prod_{m=1}^n \left(\frac{D^m g}{m!}\right)^{k_m}.\\[-1ex]
   &&\hspace{-1em}{}_{(k_1,\dots, k_n)\in P_n}\nonumber
\end{eqnarray}
To keep the notation as simple as possible, $f$ and $g$ are taken to be scalar functions, but the formula can be
generalized to $g$ being a vector and $f$ being a tensor function. The set $P_n$ denotes all $n$-tuples of nonnegative 
integers which fulfill
$1\cdot k_1+\cdots + n\cdot k_n = n$ and $D$ is the differentiation operator. Since $m$ is less or equal $n$, the 
$n$th time derivative of $f\circ g$ contains only derivatives of $g$ up to order $n$. Analogously the $n$th time derivative
of $f^{\mu\nu}(T^{\alpha0})$ contains only time derivatives of $T^{\alpha 0}$ up to order $n$.

% Non-BibTeX users please use


\begin{thebibliography}{}
%
% and use \bibitem to create references.
%
%\bibitem{RefJ}
% Format for Journal Reference
% Author, Journal \textbf{Volume}, (year) page numbers.
\bibitem{Kolb:2001qz}
  P.~F.~Kolb, U.~W.~Heinz, P.~Huovinen, K.~J.~Eskola and K.~Tuominen,
  %``Centrality dependence of multiplicity, transverse energy, and elliptic
  %flow from hydrodynamics,''
  Nucl.\ Phys.\  A {\bf 696}, (2001) 197
  [arXiv:hep-ph/0103234].

\bibitem{Shen:2011eg}
  C.~Shen, U.~Heinz, P.~Huovinen and H.~Song,
  %``Radial and elliptic flow in Pb+Pb collisions at the Large Hadron Collider
  %from viscous hydrodynamic,''
  Phys.\ Rev.\  C {\bf 84}, (2011) 044903
  [arXiv:nucl-th/1105.3226].

\bibitem{Aguiar:2001ac}
  C.~E.~Aguiar, Y.~Hama, T.~Kodama and T.~Osada,
  %``Event-by-event fluctuations in hydrodynamical description of heavy-ion collisions,''
  Nucl.\ Phys.\ A {\bf 698} (2002) 639
  [arXiv:hep-ph/0106266].

\bibitem{Qiu:2011iv}
  Z.~Qiu and U.~W.~Heinz,
  %``Event-by-event shape and flow fluctuations of relativistic heavy-ion collision fireballs,''
  Phys.\ Rev.\ C {\bf 84} (2011) 024911
  [arXiv:nucl-th/1104.0650];
%
  B.~Schenke, P.~Tribedy and R.~Venugopalan,
  %``Fluctuating Glasma initial conditions and flow in heavy-ion collisions,''
  [arXiv:nucl-th/1202.6646];
%
  B.~Schenke, S.~Jeon and C.~Gale,
  %``Elliptic and triangular flows in 3 + 1D viscous hydrodynamics with fluctuating initial conditions,''
  J.\ Phys.\ G G {\bf 38} (2011) 124169;
%
  M.~Dion, C.~Gale, S.~Jeon, J.~-F.~Paquet, B.~Schenke and C.~Young,
  %``Photons at RHIC: The Role of viscosity and of initial state fluctuations,''
  J.\ Phys.\ G G {\bf 38} (2011) 124138
  [arXiv:hep-ph/1107.0889];
%
  F.~G.~Gardim, Y.~Hama and F.~Grassi,
  %``Fluctuating Initial Conditions and Anisotropic Flows,''
  [arXiv:nucl-th/1110.5658];
%
  R.~P.~G.~Andrade, F.~Grassi, Y.~Hama and W.~-L.~Qian,
  %``Hydrodynamics: Fluctuating Initial Conditions and Two-particle Correlations,''
  Nucl.\ Phys.\ A {\bf 854} (2011) 81
  [arXiv:hep-ph/1008.0139];
%
  K.~Werner, I.~Karpenko, T.~Pierog, M.~Bleicher and K.~Mikhailov,
  %``Event-by-Event Simulation of the Three-Dimensional Hydrodynamic Evolution from Flux Tube Initial Conditions in Ultrarelativistic Heavy Ion Collisions,''
  Phys.\ Rev.\ C {\bf 82} (2010) 044904
  [arXiv:nucl-th/1004.0805];
%
  R.~Derradi de Souza, J.~Takahashi, T.~Kodama and P.~Sorensen,
  %``Effects of initial state fluctuations in the final state elliptic flow measurements using the NeXSPheRIO model,''
  Phys.\ Rev.\ C {\bf 85} (2012) 054909
  [arXiv:hep-ph/1110.5698];
%
  R.~Chatterjee, H.~Holopainen, T.~Renk and K.~J.~Eskola,
  %``Thermal photons from fluctuating initial conditions,''
  J.\ Phys.\ G G {\bf 38} (2011) 124136
  [arXiv:hep-ph/1106.3884].

\bibitem{Borsanyi:2011zzb}
  S.~Borsanyi, G.~Endrodi, Z.~Fodor, C.~Hoelbling, S.~Katz, S.~Krieg, C.~Ratti and K.~Szabo,
  %``Transition temperature and the equation of state from lattice QCD,
  %Wuppertal-Budapest results,''
  J.\ Phys.\ Conf.\ Ser.\  {\bf 316}, (2011) 012020.

\bibitem{Aoki:2005vt}
  Y.~Aoki, Z.~Fodor, S.~D.~Katz and K.~K.~Szabo,
  %``The Equation of state in lattice QCD: With physical quark masses towards the continuum limit,''
  JHEP {\bf 0601} (2006) 089
  [arXiv:hep-lat/0510084].
  %%CITATION = HEP-LAT/0510084;%%

\bibitem{Huovinen:2011xc}
  P.~Huovinen and P.~Petreczky,
  %``Equation of state at finite baryon density based on lattice QCD,''
  J.\ Phys.\ G {\bf G38} (2011) 124103
  [arXiv:nucl-th/1106.6227].
  %%CITATION = ARXIV:1106.6227;%%

\bibitem{Ejiri:2009hq}
  S.~Ejiri {\it et al.}  [WHOT-QCD Collaboration],
  %``Equation of State and Heavy-Quark Free Energy at Finite Temperature and Density in Two Flavor Lattice QCD with Wilson Quark Action,''
  Phys.\ Rev.\ D {\bf 82} (2010) 014508
  [arXiv:hep-lat/0909.2121].
  %%CITATION = ARXIV:0909.2121;%%

\bibitem{Karsch:1986cq}
  F.~Karsch and H.~W.~Wyld,
  %``Thermal Green's Functions And Transport Coefficients On The Lattice,''
  Phys.\ Rev.\ D {\bf 35} (1987) 2518.
  %%CITATION = PHRVA,D35,2518;%%

\bibitem{Aarts:2002vx}
  G.~Aarts and J.~M.~Martinez Resco,
  %``Transport coefficients from the lattice?,''
  Nucl.\ Phys.\ Proc.\ Suppl.\  {\bf 119} (2003) 505
  [arXiv:hep-lat/0209033].
  %%CITATION = HEP-LAT/0209033;%%

\bibitem{CasalderreySolana:2011us}
  J.~Casalderrey-Solana, H.~Liu, D.~Mateos, K.~Rajagopal and U.~A.~Wiedemann,
  %``Gauge/String Duality, Hot QCD and Heavy Ion Collisions,''
  [arXiv:hep-th/1101.0618].

\bibitem{Kovtun:2004de}
  P.~Kovtun, D.~T.~Son and A.~O.~Starinets,
  %``Viscosity in strongly interacting quantum field theories from black hole physics,''
  Phys.\ Rev.\ Lett.\  {\bf 94} (2005) 111601
  [arXiv:hep-th/0405231].

\bibitem{Buchel:2007mf}
  A.~Buchel,
  %``Bulk viscosity of gauge theory plasma at strong coupling,''
  Phys.\ Lett.\ B {\bf 663} (2008) 286
  [arXiv:hep-th/0708.3459].

\bibitem{Friman:2011zz}
  B.~Friman, C.~{H\"o}hne, J.~Knoll, S.~Leupold, J.~Randrup, R.~Rapp and P.~Senger (Eds.),
  %``The CBM physics book: Compressed baryonic matter in laboratory
  %experiments,''
  Lect.\ Notes Phys.\  {\bf 814}, (2011) 1.

\bibitem{Landau:1953gs}
  L.~D.~Landau,
  %``On the multiparticle production in high-energy collisions,''
  Izv.\ Akad.\ Nauk Ser.\ Fiz.\  {\bf 17}, (1953) 51.

\bibitem{Belenkij:1956cd}
  S.~Z.~Belenkij and L.~D.~Landau,
  %``Hydrodynamic theory of multiple production of particles,''
  Nuovo Cim.\ Suppl.\  {\bf 3S10}, (1956) 15.

\bibitem{Israel:1979wp}
  W.~Israel and J.~M.~Stewart,
  %``Transient relativistic thermodynamics and kinetic theory,''
  Annals Phys.\  {\bf 118}, (1979) 341.

\bibitem{Denicol:2012cn}
  G.~S.~Denicol, H.~Niemi, E.~Molnar and D.~H.~Rischke,
  %``Derivation of transient relativistic fluid dynamics from the Boltzmann
  %equation,''
  [arXiv:nucl-th/1202.4551].

\bibitem{Van:2007pw}
  P.~Van and T.~S.~Biro,
  %``Relativistic hydrodynamics - causality and stability,''
  Eur.\ Phys.\ J.\ ST {\bf 155} (2008) 201
  [arXiv:nucl-th/0704.2039].
  %%CITATION = ARXIV:0704.2039;%%

\bibitem{Van:2011yn}
  P.~Van and T.~S.~Biro,
  %``First order and stable relativistic dissipative hydrodynamics,''
  Phys.\ Lett.\ B {\bf 709} (2012) 106
  [arXiv:nucl-th/1109.0985].
  %%CITATION = ARXIV:1109.0985;%%

\bibitem{Bjorken:1982qr}
  J.~D.~Bjorken,
  %``Highly Relativistic Nucleus-Nucleus Collisions: The Central Rapidity
  %Region,''
  Phys.\ Rev.\  D {\bf 27}, (1983) 140.

\bibitem{Gubser:2010ze}
  S.~S.~Gubser,
  %``Symmetry constraints on generalizations of Bjorken flow,''
  Phys.\ Rev.\  D {\bf 82}, (2010) 085027
  [arXiv:hep-th/1006.0006].

\bibitem{Csorgo:2003rt}
  T.~Cs\"org\H o, F.~Grassi, Y.~Hama and T.~Kodama,
  %``Simple solutions of relativistic hydrodynamics for longitudinally and
  %cylindrically expanding systems,''
  Phys.\ Lett.\  B {\bf 565}, (2003) 107
  [arXiv:nucl-th/0305059].

\bibitem{Csorgo:2003ry}
  T.~Cs\"org\H o, L.~P.~Csernai, Y.~Hama and T.~Kodama,
  %``Simple solutions of relativistic hydrodynamics for systems with
  %ellipsoidal symmetry,''
  Heavy Ion Phys.\  A {\bf 21}, (2004) 73
  [arXiv:nucl-th/0306004].

\bibitem{Csorgo:2006ax}
  T.~Cs\"org\H o, M.~I.~Nagy and M.~Csanad,
  %``A New Family of Simple Solutions of Perfect Fluid Hydrodynamics,''
  Phys.\ Lett.\  B {\bf 663}, (2008) 306
  [arXiv:nucl-th/0605070].

\bibitem{Csernai:1994xw}
  L.~P.~Csernai,
  \textit{Introduction to relativistic heavy-ion collisions,}
  %\href{/spires/find/hep/www?irn=2994690}{SPIRES entry}
  (Wiley Chichester, UK 1994).

\bibitem{deHaro:2000xn}
  S.~de Haro, S.~N.~Solodukhin and K.~Skenderis,
  %``Holographic reconstruction of space-time and renormalization in the  AdS/CFT
  %correspondence,''
  Commun.\ Math.\ Phys.\  {\bf 217}, (2001) 595
  [arXiv:hep-th/0002230].

\bibitem{Janik:2010we}
  R.~A.~Janik,
  %``The dynamics of quark-gluon plasma and AdS/CFT,''
  Lect.\ Notes Phys.\  {\bf 828}, (2011) 147
  [arXiv:hep-th/1003.3291].

\bibitem{Tsumura:2012ss}
  K.~Tsumura and T.~Kunihiro,
  %``Uniqueness of Landau-Lifshitz Energy Frame in Relativistic Dissipative Hydrodynamics,''
  [arXiv:physics.flu-dyn/1206.3913].

\bibitem{Eskola:1997hz}
  K.~J.~Eskola, K.~Kajantie and P.~V.~Ruuskanen,
  %``Hydrodynamics of nuclear collisions with initial conditions from
  %perturbative QCD,''
  Eur.\ Phys.\ J.\  C {\bf 1}, (1998) 627
  [arXiv:nucl-th/9705015].

\bibitem{Shen:2010uy}
  C.~Shen, U.~Heinz, P.~Huovinen and H.~Song,
  %``Systematic parameter study of hadron spectra and elliptic flow from viscous
  %hydrodynamic simulations of Au+Au collisions at sqrt(s_NN) = 200 GeV,''
  Phys.\ Rev.\  C {\bf 82}, (2010) 054904
  [arXiv:nucl-th/1010.1856].

\bibitem{Borsanyi:2010cj}
  S.~Borsanyi, G.~Endrodi, Z.~Fodor, A.~Jakovac, S.~D.~Katz, S.~Krieg, C.~Ratti and K.~K.~Szabo,
  %``The QCD equation of state with dynamical quarks,''
  JHEP {\bf 1011} (2010) 077
  [arXiv:hep-lat/1007.2580].
  %%CITATION = ARXIV:1007.2580;%%

\bibitem{Kolb:2003dz}
  P.~F.~Kolb and U.~W.~Heinz,
  %``Hydrodynamic description of ultrarelativistic heavy-ion collisions,''
  [arXiv:nucl-th/0305084].

\bibitem{Bass:2008rv}
  S.~A.~Bass, C.~Gale, A.~Majumder, C.~Nonaka, G.~-Y.~Qin, T.~Renk and J.~Ruppert,
  %``Systematic Comparison of Jet Energy-Loss Schemes in a realistic hydrodynamic medium,''
  Phys.\ Rev.\ C {\bf 79} (2009) 024901
  [arXiv:nucl-th/0808.0908].

\bibitem{Rapp:2011is}
  R.~Rapp,
  %``Theory of Soft Electromagnetic Emission in Heavy-Ion Collisions,''
  Acta Phys.\ Polon.\ B {\bf 42} (2011) 2823
  [arXiv:nucl-th/1110.4345].
  %%CITATION = ARXIV:1110.4345;%%

\bibitem{Knuth:73}
  D.~E.~Knuth,
  \textit{The Art of Computer Programming Vol. 1 - Fundamental Algorithms}
  (Addison-Wesley Publishing Company, Reading, Massachusetts 1973) 50.

% Format for books
%\bibitem{RefB}
%Author, \textit{Book title} (Publisher, place year) page numbers
% etc


\end{thebibliography}
\end{document}